**RESEARCH ARTICLE**

# Mediastinal Lymph Node Detection and Segmentation Using Deep Learning

AL-AKHIR NAYAN[1], BOONSERM KIJSIRIKUL[1], AND YUJI IWAHORI[2], (Member, IEEE)
[1] Department of Computer Engineering, Faculty of Engineering, Chulalongkorn University, Bangkok 10330, Thailand
[2] Department of Computer Science, College of Engineering, Chubu University, Kasugai, Aichi 487-8501, Japan

Corresponding author: Al-Akhir Nayan (asquiren@gmail.com)

This work was supported by the Japan Society for the Promotion of Science (JSPS) Grant-in-Aid Scientific Research (C) under Yuji Iwahori's Grant number 20K11873. Al-Akhir Nayan was also supported by Chulalongkorn University's Graduate Scholarship Programme for ASEAN or Non-ASEAN Countries.

**ABSTRACT** Automatic lymph node (LN) segmentation and detection for cancer staging are critical. In clinical practice, computed tomography (CT) and positron emission tomography (PET) imaging detect abnormal LNs. Despite its low contrast and variety in nodal size and form, LN segmentation remains a challenging task. Deep convolutional neural networks frequently segment items in medical photographs. Most state-of-the-art techniques destroy image's resolution through pooling and convolution. As a result, the models provide unsatisfactory results. Keeping the issues in mind, a well-established deep learning technique UNet++ was modified using bilinear interpolation and total generalized variation (TGV) based upsampling strategy to segment and detect mediastinal lymph nodes. The modified UNet++ maintains texture discontinuities, selects noisy areas, searches appropriate balance points through backpropagation, and recreates image resolution. Collecting CT image data from TCIA, 5-patients, and ELCAP public dataset, a dataset was prepared with the help of experienced medical experts. The UNet++ was trained using those datasets, and three different data combinations were utilized for testing. Utilizing the proposed approach, the model achieved 94.8% accuracy, 91.9% Jaccard, 94.1% recall, and 93.1% precision on COMBO_3. The performance was measured on different datasets and compared with state-of-the-art approaches. The UNet++ model with hybridized strategy performed better than others.

## I. INTRODUCTION

Cancer has the highest death rate, with 59.9 million dying cases yearly [1]. Cancer patient's survival rates are decreasing over time. According to a UK Cancer Research Centre study, roughly 30% of patients will survive if they have been sick for a year, but just 5% would survive if they have been sick for ten years [2]. As a result, early detection, diagnosis, and treatment are critical. Furthermore, with the advancement of radiology in recent years, CT scans have aided cancer trials. Lymph nodes (LNs) play a crucial role in cancer treatment and help determine therapy response [3]. LNs segmentation aims to give a categorical label to each pixel in an image. However, their oblique border and low contrast with adjacent tissue and organs on CT images make LNs segmentation and detection difficult [4]. As a result, manually counting and quantifying LNs in images from human observers is time-consuming and error-prone [5].

Deep Learning algorithms have lately gained popularity in medical image analysis. It performs better than traditional statistical and atlas-based machine learning techniques and maintains good speed and accuracy [6]. Applications using fully convolutional networks (FCNs) gained popularity in medical image segmentation [7]. Following FCNs, various alternative convolutional neural network-based segmentation architectures, such as SegNet [8], U-Net [9], and DeepLab [10], have been proposed. However, few articles on deep learning based LNs segmentation have been published. FCNs were trained to learn the appearance of lymph node clusters or contour probabilistic output maps in [11], and

  89289

conditional random fields were employed to segment LNs. To control the balance of sizes between the target classes in [12], a 3D U-Net was used to partition mediastinal LNs and other anatomical features like lungs, airways, aortic arches, and pulmonary arteries.

However, deep learning algorithms in the medical image segmentation sector encounter several roadblocks. It is relatively costly to create a large dataset (such as ImageNet [13]) for medical images because it requires radiologists to name the anatomical objects in the images. Besides, some minor or unusual observations will result in label imbalance when the deep segmental neural network is trained. Previous studies on mediastinal LN segmentation show several issues due to the lack of necessary datasets and excellent segmentation strategy. Besides, most studies have overlooked information loss issues caused by the max pool and convolution process and given minor importance to the small but substantial portion of medical images. Such ignorance has led those models to show unsatisfactory performance.

In this research, the drawbacks of previous approaches have been overcome concerning new strategies. For better training, a dataset was prepared with the help of medical specialists with a background of several years in the related field. A trainable upsampling strategy was introduced with the UNet++ model for recreating the image resolution destroyed by the max pooling and convolution process. The strategy confirms an information lossless training procedure for the neurons. Different combinations of several datasets, such as TCIA, 5-patients, and the ELCAP public dataset, were employed to examine the segmentation and detection ability of the hybridized UNet++ model. The capability of the proposed technique was compared with state-of-the-art approaches such as Auto_LNDS, SegNet, and AlexNet. Accuracy, precision, dice score, recall, and F1 score were parameters for finding the best approach.

The article is organized as follows: related studies, their success, and drawbacks have been explained in section 2. Section 3 describes dataset pre-processing, modified model's architecture, proposed upsampling strategy, and evaluation process. The model's performance on different datasets and the comparison among several state-of-the-art approaches have been mentioned in section 4. Research limitations have been discussed in section 5. Lastly, the whole work has been summarized in section 6.

## II. RELATED WORKS

Most previous studies on computer-aided detection (CADe) systems for LNs depend on direct 3D data from volumetric CT images. Barbu *et al.* employed boosting-based feature extraction and integration to obtain a reliable binary classifier on preferred features. On 131 volumes having 371 LNs, the approach was assessed for axillary LN identification, yielding an 83.0% detection rate with 1.0 False Positive (FP)/volume. However, the number of training specimens was limited, and 3D CT scans had a heightened dimensionality [14].

Oda *et al.* introduced a fully convolutional network-based technique using 3D U-Net for detecting and segmenting mediastinal LNs from chest CT volumes. The testing results revealed that 95.5% of lymph nodes were identified with 16.3 FP/CT volume. Still, the model is likely to be overfitted due to many parameters. Solving the over-segmentation and overfitting issues, the model gained 82% accuracy. Still, several regions of the chest anatomies should be incorporated into the dataset to negotiate the size inequality between classes [12].

Bouget *et al.* suggested that a technique for precise and automatic segmentation is essential for describing lymph nodes numerically [15]. Their study used downsampled full volumes or slab-wise techniques to create 3D convolutional neural networks. Neighboring anatomical features such as similar attenuation values of lymph nodes were used as prior knowledge to segment the tissue. A dataset of 120 contrast-enhanced CT volumes and a 5-fold cross-validation approach was used to evaluate the performances. The method achieved a patient-wise recall of 92% with a false positive per patient ratio of 5 and a segmentation overlap of 80.5% for the 1178 lymph nodes with 10 mm diameter. The best results were combined with a slab-wise and a full volume approach. However, more than four organs were required to produce the best results, which indicated the need for a large dataset.

To identify abdominal LNs (ALNs) and mediastinal LNs (MLNs) in CT images, Tekchandani *et al.* suggested attention U-Net based deep learning architecture variants with complete and partial transfer learning [16]. The attention U-Net was chosen because it could concentrate more on target structures with trainable parameters. The full and partial transfer learning (TL) tackled the limited image dataset. The attention U-Net was trained using ALNs, and MLNs CT images, and several experiments were conducted to determine the usefulness of the suggested methodology. The proposed strategy was compared with methods such as SegNet, U-Net, and ResUNet. The highest achieved sensitivity and Dice scores for ALNs were 91.69% and 93.08%, but the model seemed overfitted in segmenting MLNs.

Seff *et al.* [17] described a solution for separating 3D identification issues into 2D detection subtasks on CT scans by separating a prospect into 27 CT slices, which alleviated the scourge of dimensionality problems to a great extent. They used the Histogram of Oriented Gradients (HOG) to extract features and then performed linear classification. The mediastinal and abdominal datasets reached 78.0% sensitivity at 6 FP/vol. (86.1% at 10 FP/vol.) and 73.1% sensitivity at 6 FP/vol. (87.2% at 10 FP/vol.) respectively. They also used baseline HOG systems and a state-of-the-art deep CNN [18] to detect mediastinal LNs, with sensitivities of 78% vs. 70% at 3 FP/scan and 88% vs. 84% at 6 FP/scan. However, a more efficient strategy was required to enhance accuracy and sensitivity.

Roth *et al.* proposed a 2.5D detecting approach to amplify the samples. Their methods rapidly decomposed each 3D volume of interest (VOI) by resampling 2D reformatted orthogonal views. The CNN classifier was trained using those random views. Their experiment achieved 70% / 83% sensitivity at 3 FP/vol. and 84% / 90% sensitivity at 6 FP/vol [19].

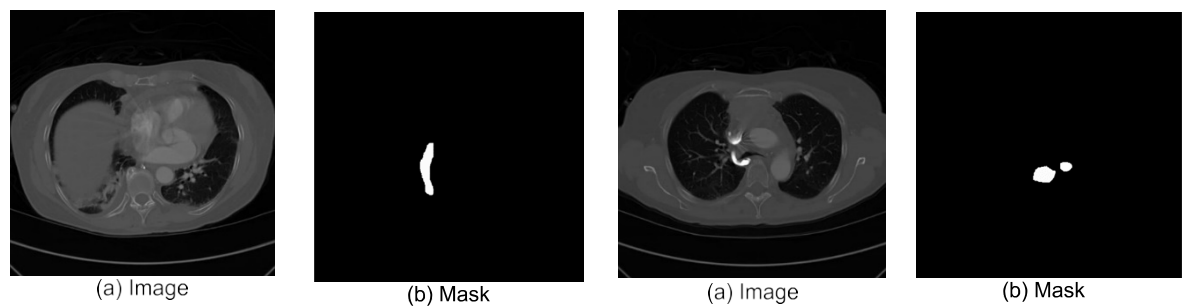

**FIGURE 1.** Dataset sample a) Image, b) Mask.

To accomplish the detection and segmentation of the malignant lymph nodes, Zhu *et al.* suggested a multi-branch detection by segmentation network. An effective distance-based gating technique was established in their suggested framework, mimicking protocols employed by oncologists in everyday practice. Unfortunately, a PET/CT scan was typically obtained later in the diagnostic process. As a result, before assuming the PET/CT modality, optimum processing of the first CT volume was required [20].

Most of the studies for LN segmentation have shown several issues and performed poor detection accuracy. That issues declare the need for an efficient approach to improve the accuracy and sensitivity of LN segmentation. Through this research, we have tried to implement a technique that performs better than other approaches.

## III. RESEARCH METHOD
### A. DATASET
The TCIA [21], 5-patients [22], and the ELCAP [23] public datasets were obtained, processed, and utilized as the training dataset in this work. The testing dataset was created, combining data from those three datasets. Some sample images and their masks are shown in Figure 1.

Three radiologists of various seniority (21, 20, and 30 years, respectively) recommended the mentioned mediastinal LN datasets to conduct this research. All LNs were masked using VGG Image Annotation Toolkit by two veteran radiologists with 21 and 20 years of expertise. If there was a discrepancy, a third senior radiologist (30 years of expertise) was called in to decide about the presence of LN. In total, 28830 LNs were masked in the training dataset, which was used to train the UNet+ model. Another three testing datasets, each with an average of 8500 images, were segmented for testing and evaluating the model's performance.

### B. DATA PROCESSING
High-quality CT images were used in this investigation because the higher resolution generates the most substantial diffusion effects. The background tissue signal suppresses well on high-resolution CT images and allows the high signal intensity LNs to be exhibited and identified. For segmentation, images were manually cropped with a 512$\times$512 matrix. Finally, 28830 processed images were used as the training dataset, and another three datasets, each with 8500 images, served as the testing dataset.

The model was trained using the TCIA, 5-patients, and ELCAP public datasets to investigate its capability on different datasets. For testing purposes, there were three combinations. 80% of data from TCIA, 10% from the 5-patients, and 10% from the ELCAP public dataset were tried as the first combination (COMBO_1). 60% from TCIA, 20% from the 5-patients, and 20% from the ELCAP public dataset were utilized as the second combination (COMBO_2). 50% from TCIA, 25% from the 5-patients, and 25% from the ELCAP public dataset were tested as the third combination (COMBO_3). The training dataset was split into two parts (train and validation) using the Python data split library.

### C. DATA AUGMENTATION
In the context of CNN, artificial data augmentation is a standard approach for producing sufficient training data. When the dataset is insufficient, it can also teach the network the desired invariances and resilience qualities [24]. The Python data augmentation module enhanced the training dataset in this work. Image cropping, affine modifications, vertical and horizontal flipping, noise and blur reduction, and contrast and brightness controlling were performed. The training dataset was augmented in this work, but the testing data was kept un-augmented.

### D. MODIFIED MODEL ARCHITECTURE
A modified UNet+ network was adapted to find Lymph Node candidates from mediastinal CT volumes. The UNet+ is a convolutional network based on U-Net for segmenting biological images quickly and precisely. The network can be separated into an encoder-decoder path or a contracting-expansive path. The encoder is made up of two 3$\times$3 convolutions that are applied repeatedly.

In our work, there was a ReLU and batch normalization after each convolution. Batch normalization reduced covariate shift, allowing each layer to learn more independently of the others. The spatial dimensions were reduced using a 2 ×2 max pooling operation. A dropout layer before max-pooling functions as a regularizer and helps to avoid overfitting. The number of feature channels was doubled while half of the spatial dimensions were cut at each down-sampling step. Every step in the expanding route started with an upsampling of the feature map. Bilinear interpolation and total generalized variation (TGV) were utilized to maintain texture discontinuities, select noisy areas, search for appropriate balance points through backpropagation, and recreate image resolution. To bridge the semantic gap between the feature maps of the encoder and decoder, nested and dense skip connections [25] were utilized before concatenation. The Modified UNet++ model architecture is shown in Figure 2, where thick convolution blocks on the skip paths are green and blue. We had a 3 × 3 convolutional concatenation with the equivalent feature map from the contracting path. A 1×1 convolution maps each 16-component feature vector and ensures appropriate image segmentation at the final layer.

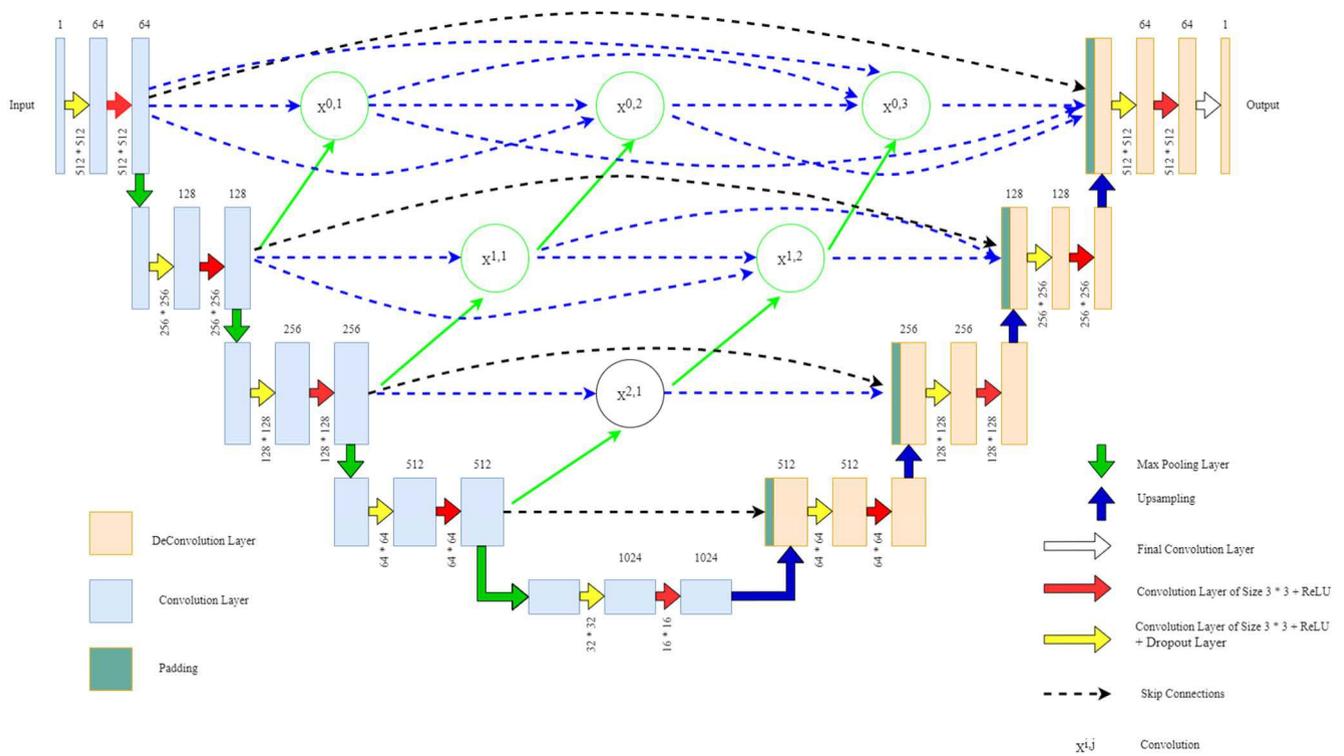

**FIGURE 2.** Modified UNet++ Model architecture.

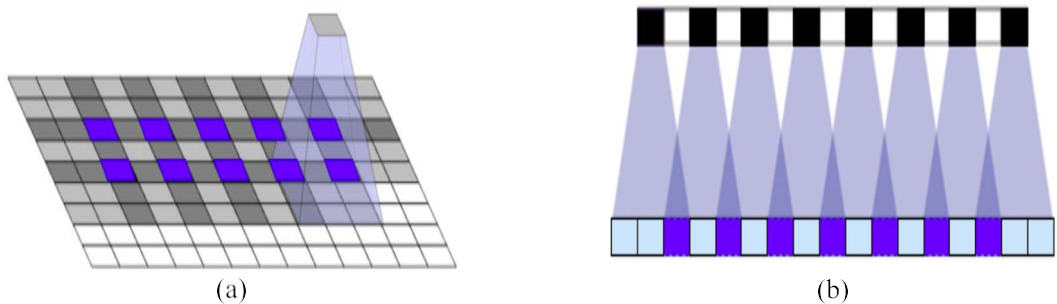

**FIGURE 3.** Artifact issue created by the transposed convolution: (a) checkerboard problem, (b) uneven overlap.

### E. UPSAMPLING TECHNIQUE

The spatial information of the feature map is generated through upsampling; that is why it is the most significant aspect of the UNet architecture. By default, the network utilizes transpose convolution in the upsampling section. Still, this execution approach takes a long time to complete because of having a massive number of learnable parameters and the need to train additional weights by kernels. Furthermore, it can result in unequal overlap and artifacts at various scales, as shown in Figure 3.

An alternative to the traditional deconvolution, bilinear interpolation was employed in the decoder section of the model to avoid artifacts. This approach has no artifacts because of its normal behavior. The image's $g(x_1, x_2)$ nearest neighbors are the image coordinates $f(x_{10}, x_{20})$, $f(x_{11}, x_{21})$, $f(x_{12}, x_{22})$ and $f(x_{13}, x_{23})$. The interpolated image's $g(x_1, x_2)$ intensity values are evaluated as follows:

$$g(x_1, x_2) = B_0 + B_1 x_1 + B_2 x_2 + B_3 x_1 x_2 \qquad (1)$$

Equation (1) denotes a bilinear function that has coordinates $(x_1, x_2)$. Here, $g$ refers to an intensity value of interpolated image matrix in $(x_1, x_2)$ coordinates, Bs are the bilinear weights, and f refers intensity value given at $(x_{10}, x_{20})$, $(x_{11}, x_{21})$, $(x_{12}, x_{22})$ and $(x_{13}, x_{23})$ pixel locations before performing interpolation. The bilinear weights, $B_0$, $B_1$, $B_2$ and $B_3$ are calculated by evaluating the matrix mentioned in Equation (2):

$$\begin{bmatrix} B_0 \\ B_1 \\ B_2 \\ B_3 \end{bmatrix} = \begin{bmatrix} 1 & x_{10} & x_{20} & x_{10}x_{20} \\ 1 & x_{11} & x_{21} & x_{11}x_{21} \\ 1 & x_{12} & x_{22} & x_{12}x_{22} \\ 1 & x_{13} & x_{23} & x_{13}x_{23} \end{bmatrix} \qquad (2)$$

As an outcome, $g(x_1, x_2)$ is determined as a linear combination of its four closest neighbor's gray levels. According to Equation (1), when the perfect least-squares planar fit is made to those four neighbors, the value is assigned to $g(x_1, x_2)$. This optimal averaging methodology provides smoother outcomes considering the low-and high-resolution map's size difference. The bilinear interpolation technique is depicted in Figure 4.

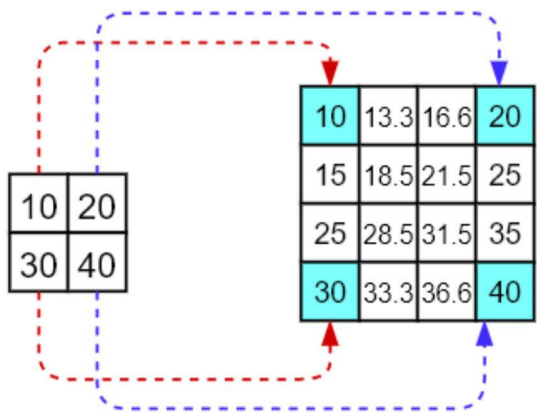

**FIGURE 4.** Example of bilinear interpolation.

The result generated by bilinear interpolation contains the equal size as processed maps in the encoding stage. These maps refer to the entered image and are seen as noisy areas. The TGV treats the processed maps for denoising resolution. The mathematical term used for the model is mentioned in Equation (3):

$$\min_u \{\gamma \, loss(u, x) + TGV_\alpha^K(v)\} \qquad (3)$$

Here, image fidelity is represented by $loss(u, x)$, $\gamma$ is utilized for global optimization, and the term $TGV_\alpha^K(v)$ is the regularization term.

Convolutional computations may effectively extract and generalize features, and their outputs can be visualized as a set of features. To better comprehend the semantics of an image, the smoothness of textures and border edges are essentially used for the feature maps. The mentioned technique maintains texture discontinuities by incorporating different channel information.

For our work, a second-order TGV-based technique was enough. That is why the utilized Equation (4) was:

$$TGV_a^2(v) = \min_w \, p_1 \, |\nabla v - w| + p_2 \, |\varepsilon(w)| \qquad (4)$$

Here, v indicates the minimum TGV over complex vector fields w in the bounded domain. The symmetric derivative is represented by $\varepsilon(w)$. $p_1$ and $p_2$ are responsible for balancing the first and second derivatives.

Combining loss function and Equation (3), the term TGV can be defined as

$$\min_w \begin{Bmatrix} \lambda \| u_{C_{input}} - C_{input} \|_2 + \\ \frac{2}{p} \, |\nabla v - w| + p_2 \, |\varepsilon(w)| \end{Bmatrix} \qquad (5)$$

Taking weights $p_1$ and $p_2$ into backpropagation, an appropriate balance point was decided by the network.

### F. BACKPROPAGATION

Backpropagation is a supervised algorithm used for adjusting the weights $p_1$, $p_2$ and the network's deviation. As a result, the output and the targeted vectors can be closer. It compares the error sum, and the training is finished if the network's output layer's error is evaluated as lower than the specified error. The network's deviation and weight are stored. The process is performed following the steps:

1. Setting nonzero coefficient and initial weight $p_{ik}$ but $p_{i,n+1} = -threshold$
2. Entering sample $A = (A_1, A_2, \ldots)$ and getting the desired output $B = (B_1, B_2, \ldots)$.
3. Calculating output of all layers. For output $A_i^j$ there is:

$$X_i^j = \sum_{k=1}^{m+1} p_{ik} A_k^{j-1} \qquad (6)$$

$$A_i^j = f(X_i^q) \qquad (7)$$

4. Finding learning errors for each layer. For the output layer, j=n:

$$e_i^n = A_i^n (1 - A_i^n)(A_i^n - B_i) \qquad (8)$$

5. Modify weight and threshold.

$$t\text{-}, p_{ik}(h + 1) = t\text{-}, p_{ik}(h) - \eta . e_i^n . A_k^{j-1} \qquad (9)$$

6. Ensure the quality of weights according to the requirement. If the requirement is not filled up, return to step 3. The process will be repeated until the requirement is satisfied.

### G. EVALUATION PROCESS

The following approach was used to calculate the parameter's value: True Positives (TP) are accurately predicted positive values, indicating that the value of the actual class and the value of the projected class are both yes. True Negatives (TN) are accurately predicted negative values, implying that the value of the actual class is zero and the value of the projected class is also zero. False positives and false negatives are values that arise when the actual class and the projected class are not the same. When the actual class is no and the projected class is yes, False Positives (FP) occurs. When the actual class is yes, but the projected class is no, this is known as False Negatives (FN). Precision, recall, F1, and accuracy were evaluated to assess the model's performance. The higher value indicates better performance. The most straightforward intuitive performance metric is accuracy, which is the ratio of correctly predicted and all observations. The model's accuracy was measured utilizing the formula mentioned in Equation (10).

$$Accuracy = \frac{TP + TN}{TP + FP + FN + TN} \quad (10)$$

The ratio of accurately predicted positive observations and total expected positive observations is precision. The precision of the model was calculated following Equation (11).

$$Precision = \frac{TP}{TP + FP} \quad (11)$$

The ratio of accurately predicted positive observations and all actual yes class observations is known as recall. Precision and Recall are weighted into the F1 Score. As a result, false positives and negatives are considered while calculating this score. When there is an unequal distribution of classes, F1 is usually more helpful. Recall and F1 are calculated following Equations (12) and (13) accordingly.

$$Recall = \frac{TP}{TP + FN} \quad (12)$$

$$F1 = \frac{2 * Recall * Precision}{Recall + Precision} \quad (13)$$

### H. LYMPH NODE SEGMENTATION CRITERIA

The Dice similarity coefficient (DSC) measures the similarity between the segmentation results and the ground truth. The DSC is a number from 0 to 1, with a more excellent value indicating better segmentation accuracy. The DSC was established as follows:

$$DSC(R, T) = \frac{2P(R \cap T)}{P(R) + P(T)} \quad (14)$$

Here R is the segmentation result of the segmentation method, T is the ground truth, and P is the number of pixels in the corresponding set.

## IV. RESULT AND ANALYSIS
### A. EXPERIMENTAL ENVIRONMENT SETUP

The UNet+ model was trained on an Ubuntu 18.04 GPU server with 32 Intel CPUs having a Tesla 16 Gb GPU and 64 GB accessible in RAM. The layers were trained using an Adam Optimizer with the default value (b1= 0.9, b2 =0.999) and 32 images per batches. Table 1 illustrates the training hyperparameters. 10% of the 28830 images in the training dataset were utilized to evaluate the deep-learning model's learning effect (as a validation set), while the rest were employed to train the model.

**TABLE 1.** Model parameters.

| Parameter Name | Value |
|---|---|
| Epochs | 100 |
| Learning Rate | 1.e - 4 |
| Batch Size | 32 |
| Optimizer | Adam |
| Input Image | 512 * 512 |

This study employed a 10-fold cross-validation method to validate CT data. A cross-validation run entails dividing the entire dataset into subgroups; the training set and the validation set or testing set. Several cross-validations with different random partitions were frequently executed. Finally, the model performance was estimated using the average validation data acquired from several rounds of cross-validation. The model's loss changes were checked on the validation set during the training phase to make alternative decisions on the learning rate and other parameters. The learning rate would be reduced to half of its present value if the loss of the validation set did not decrease for ten epochs. To avoid over-fitting, the training procedure must be stopped early if the loss of the validation set did not reduce for 20 epochs. Finally, the network's output was a probability map for the background and foreground. Primary data augmentation was added to the train set, leaving the test set unchanged. The final prediction results were evaluated by averaging the model's findings on the enriched data.

### B. PERFORMANCE MEASUREMENT ON TCIA DATASET
#### 1) TRAINING ACCURACY MEASUREMENT

Three radiologists with diverse expertise (21, 20, and 30 years, respectively) in mediastinal LNs imaging diagnosis rechecked the CT images before training. After rechecking, the modified UNet model was trained using 34 mediastinal LN slices from the TCIA dataset; each contained approximately 600 images on average. 10% of images from the training set were employed for tuning the model's learning performance. The model was trained for 100 epochs, and the computed parameter's value is mentioned in Figure 5.

Figure 5 (a) depicts the model's training accuracy, and loss on the TCIA dataset, 5 (b) represents validation accuracy and loss, 5 (c) illustrates intersect of union (IOU) score, 5 (d) displays dice coefficient score, and 5 (e) mentions the score of recall. The model carried 97.1% training accuracy with 93.4% validation. The training and validation loss was tremendous at the beginning of the training process. The losses were minimized quickly with the increasing number of epochs concluding that the model could learn faster. The IOU, dice coefficient, and recall scores were 90.8%, 92.5%, and 94.4%.

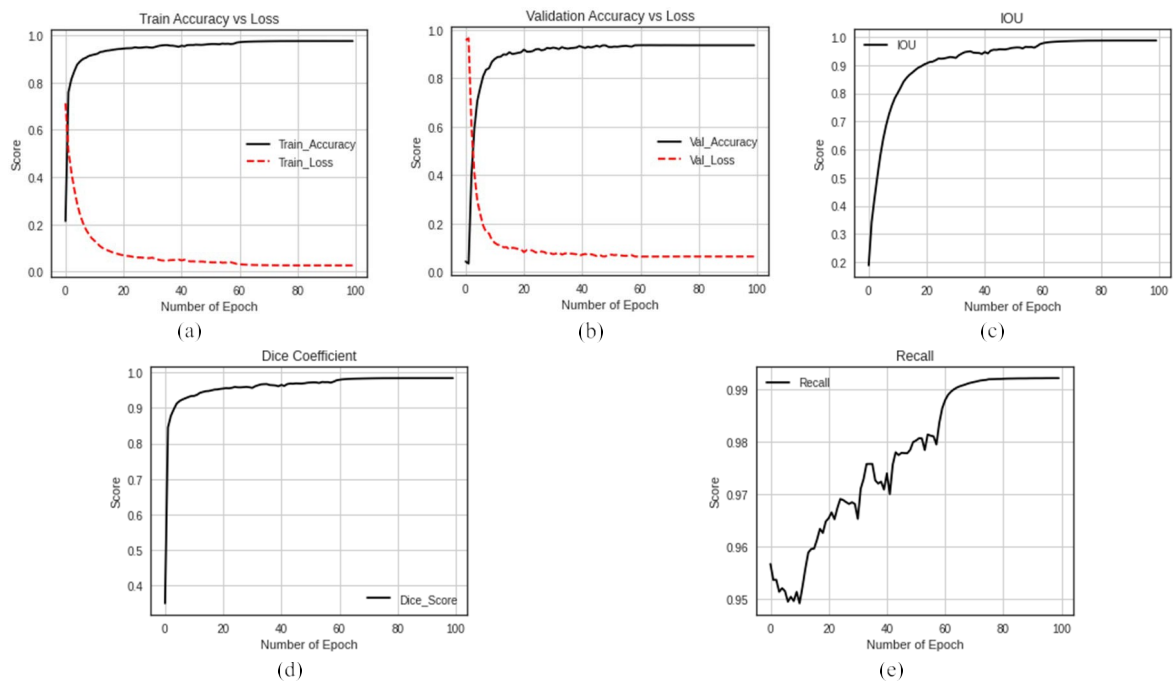

**FIGURE 5.** Training score evaluation a) Train Accuracy vs. Loss, b) Validation Accuracy vs. Loss, c) Intersect of Union (IOU), d) Dice Coefficient, e) Recall.

#### 2) TESTING ACCURACY MEASUREMENT
We assembled three testing datasets integrating images from TCIA, ELCAP, and 5-patients dataset to investigate the model's accuracy on different datasets. Each variety (COMBO_1, COMBO_2, and COMBO_3) contained 8500 mediastinal lymph node CT images and their corresponding masks. The testing datasets were constructed and inspected by radiologists with experience of several years in related fields.

*a: EXPERIMENTAL ANALYSIS ON COMBO_1*
Firstly, the model's performance was calculated on COMBO_1. The combination was created, gathering 80% of images from TCIA, 10% from the 5-patients, and 10% from the ELCAP public dataset. The performance was assessed using five parameters: accuracy, Jaccard, precision, recall, and F1. Figure 6 depicts the model's performance on the COMBO_1 dataset.

Figure 6 (a) shows the value of the accuracy parameter, 6 (b) depicts the Jaccard score, 6 (c) illustrates precision, 6 (d) demonstrates recall, and 6 (e) displays the F1 parameter's score. The model attained 91.1% accuracy, 89.0% Jaccard, 87.4% precision, 89.7% recall and 88.5% F1 accordingly.

*b: EXPERIMENTAL ANALYSIS ON COMBO_2*
The COMBO_2 was prepared, gathering 60% images from TCIA, 20% from the 5-patients, and 20% from the ELCAP public dataset. Figure 7 shows the model's performance on the COMBO_2 dataset.

Figure 7 (a) depicts the value of the accuracy parameter, 7 (b) shows the Jaccard score, 7 (c) represents precision, 7 (d) shows recall, and 7 (e) displays the F1 parameter's score. The model achieved 92.7% accuracy, 90.5% Jaccard, 89.3% precision, 91.4% recall, and 90.3% F1.

*c: EXPERIMENTAL ANALYSIS ON COMBO_3*
The COMBO_3 was prepared, assembling 50% images from TCIA, 25% from the 5-patients, and 25% from the ELCAP public dataset. Figure 8 illustrates the model's performance on the COMBO_3 dataset.

Figure 8 (a) depicts the value of the accuracy parameter, 8 (b) illustrates the Jaccard score, 8 (c) represents precision, 8 (d) shows recall, and 8 (e) displays the F1 parameter's score. The model earned 93.8% accuracy, 91.7% Jaccard, 89.8% precision, 93.2% recall, and 91.5% F1. Using COMBO_3 better performance was achieved.

### C. PERFORMANCE MEASUREMENT ON ELCAP DATASET
#### 1) TRAINING ACCURACY MEASUREMENT
The modified UNet model was trained using 20 mediastinal LN slices from the ELCAP dataset; each possessed approximately 250 images on average. 10% of images from the training set were utilized for tuning the model's learning performance. The model was trained for 100 epochs, and the computed parameter's value is mentioned in Figure 9.

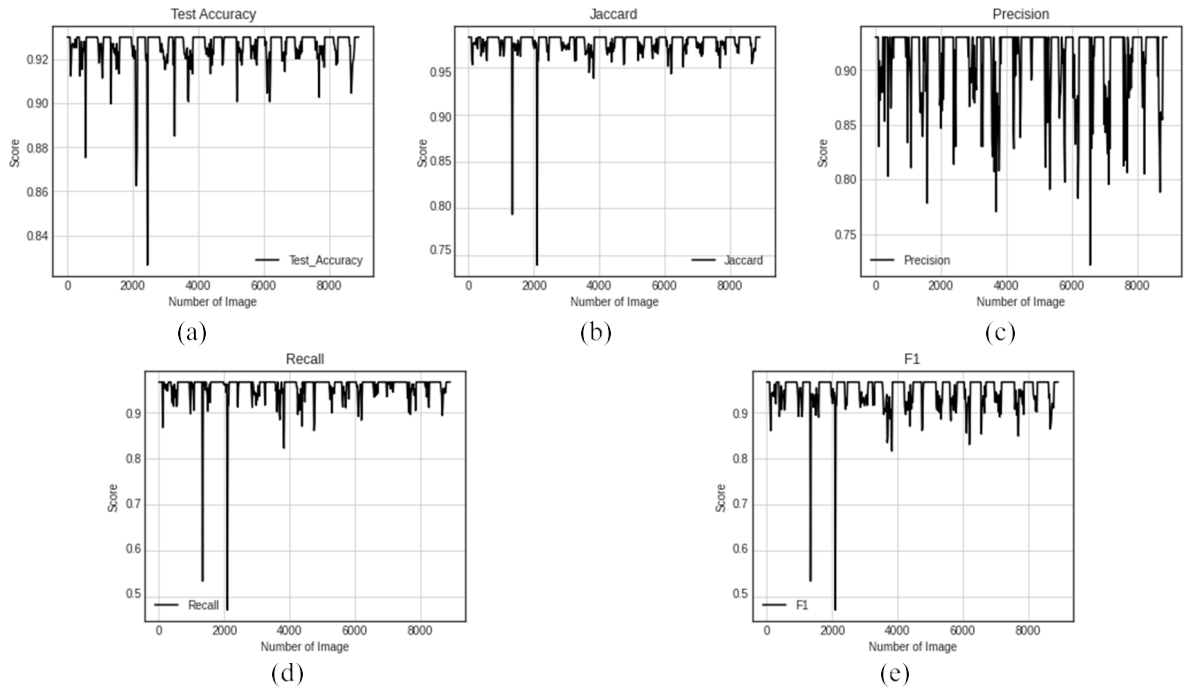

**FIGURE 6.** Testing accuracy measurement on test dataset COMBO_1, a) Accuracy, b) Jaccard, c) Precision, d) Recall, e) F1.

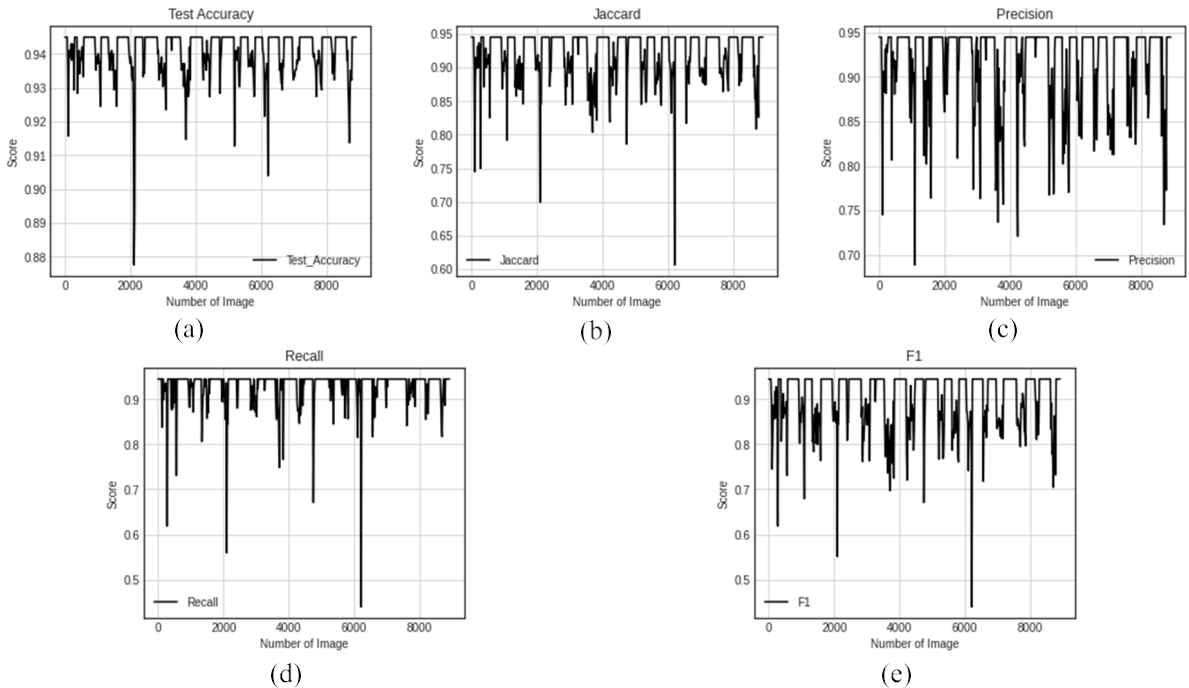

**FIGURE 7.** Testing accuracy measurement on test dataset COMBO_2, a) Accuracy, b) Jaccard, c) Precision, d) Recall, e) F1.

Figure 9 (a) displays the model's training accuracy and loss on the ELCAP dataset, 9 (b) represents validation accuracy, and loss, 9 (c) shows intersect of union (IOU) score,

9 (d) displays dice coefficient score, and 9 (e) mentions the score of recall. The model carried 97.1% training accuracy with 95.04% validation. The training and validation loss was tremendous at the beginning of the training process. The losses were minimized quickly with the increasing number of epochs concluding that the model could learn faster. The IOU,

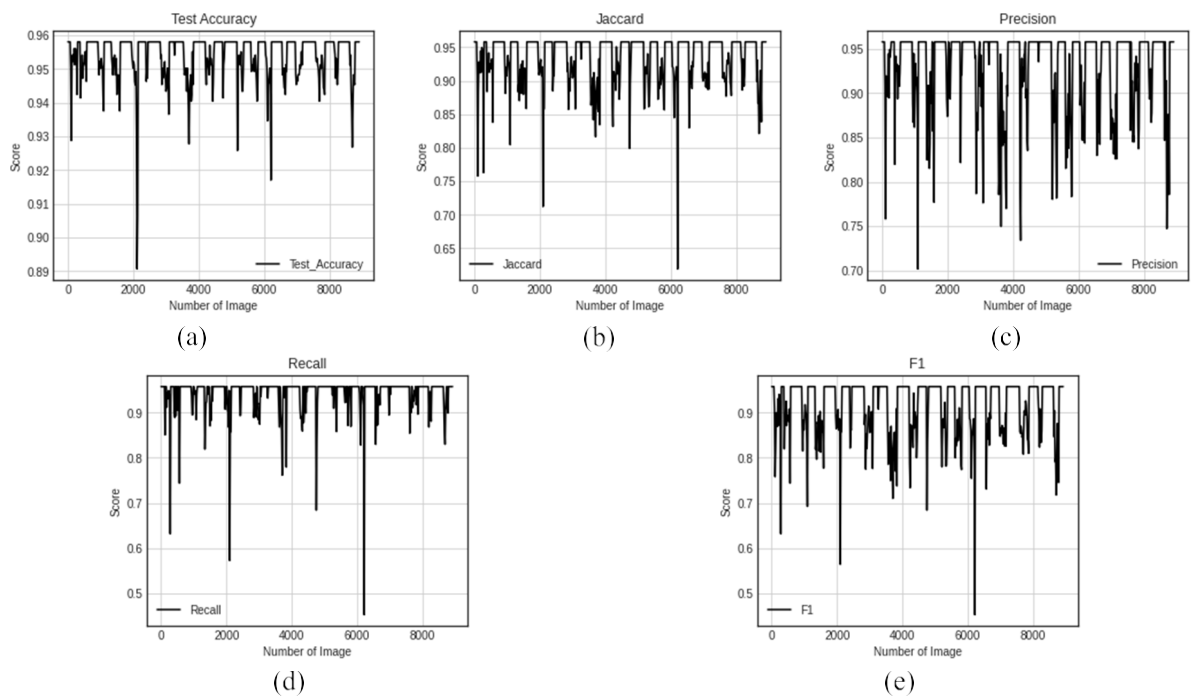

**FIGURE 8.** Testing accuracy measurement on test dataset COMBO_3, a) Accuracy, b) Jaccard, c) Precision, d) Recall, e) F1.

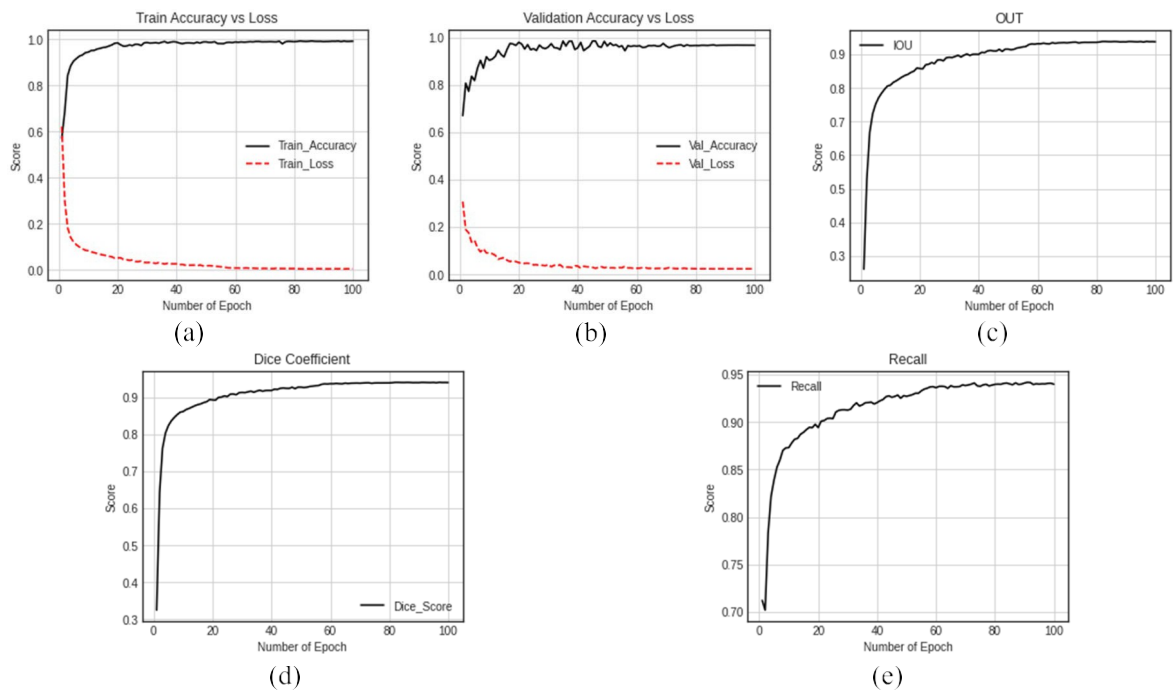

**FIGURE 9.** Training score evaluation a) Train Accuracy vs. Loss, b) Validation Accuracy vs. Loss, c) Intersect of Union (IOU), d) Dice Coefficient, e) Recall.

dice coefficient and recall score were found 89.8%, 91.5%, and 93.8%.

#### 2) TESTING ACCURACY MEASUREMENT
*a: EXPERIMENTAL ANALYSIS ON COMBO_1*
Figure 10 (a) illustrates the value of the accuracy parameter, 10 (b) shows the Jaccard score, 10 (c) depicts precision, 10 (d) exhibits recall, and 10 (e) displays the F1 parameter's score. The model achieved 92.1% accuracy, 89.6% Jaccard, 88.2% precision, 91.8% recall, and 89.9% F1.

*b: EXPERIMENTAL ANALYSIS ON COMBO_2*
Figure 11 (a) depicts the value of the accuracy parameter, 11 (b) displays the Jaccard score, 11 (c) depicts precision,

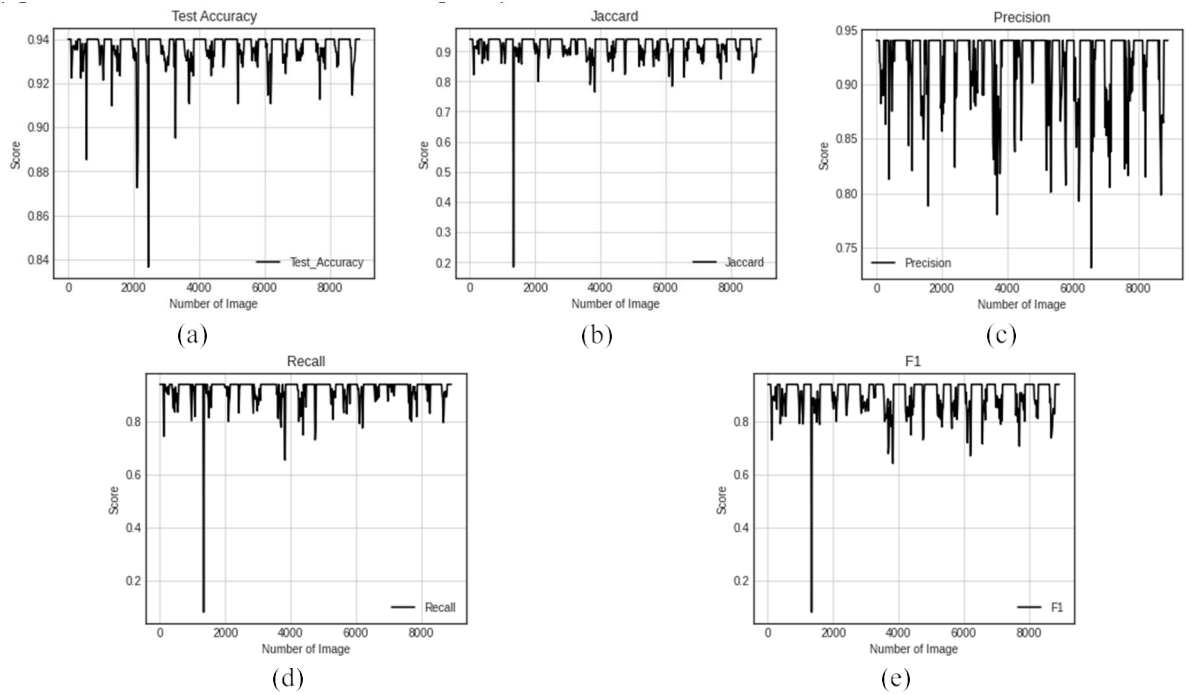

**FIGURE 10.** Testing accuracy measurement on Test Dataset COMBO_1, a) Accuracy, b) Jaccard, c) Precision, d) Recall, e) F1.

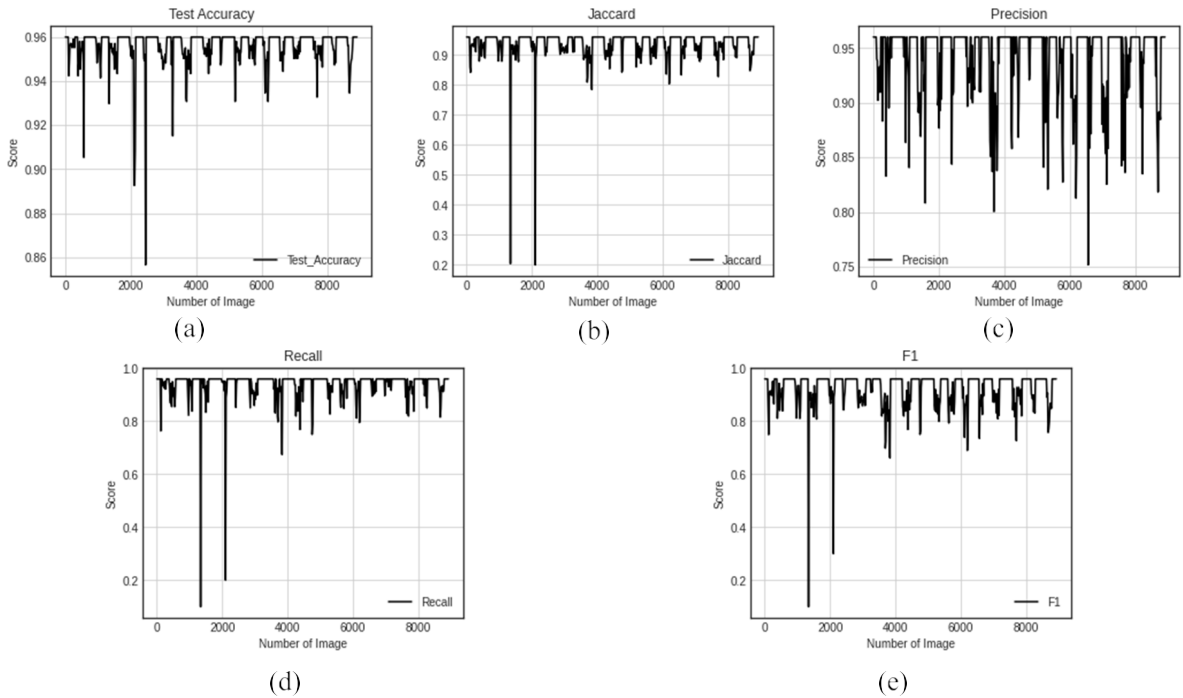

**FIGURE 11.** Testing accuracy measurement on test dataset COMBO_2, a) Accuracy, b) Jaccard, c) Precision, d) Recall, e) F1.

11 (d) shows recall, and 11 (e) illustrates the F1 parameter's score. The model attained 94.1% accuracy, 93.6% Jaccard, 90.2% precision, 93.8% recall, and 91.9% F1. Utilizing COMBO_2, satisfactory performance was achieved.

*c: EXPERIMENTAL ANALYSIS ON COMBO_3*
Figure 12 (a) shows the value of the accuracy parameter, 12 (b) illustrates the Jaccard score, 12 (c) depicts precision, 12 (d) shows recall, and 12 (e) displays the F1 parameter's score. The model acquired 93.6% accuracy, 91.7% Jaccard, 90.1% precision, 92.9% recall, and 91.5% F1.

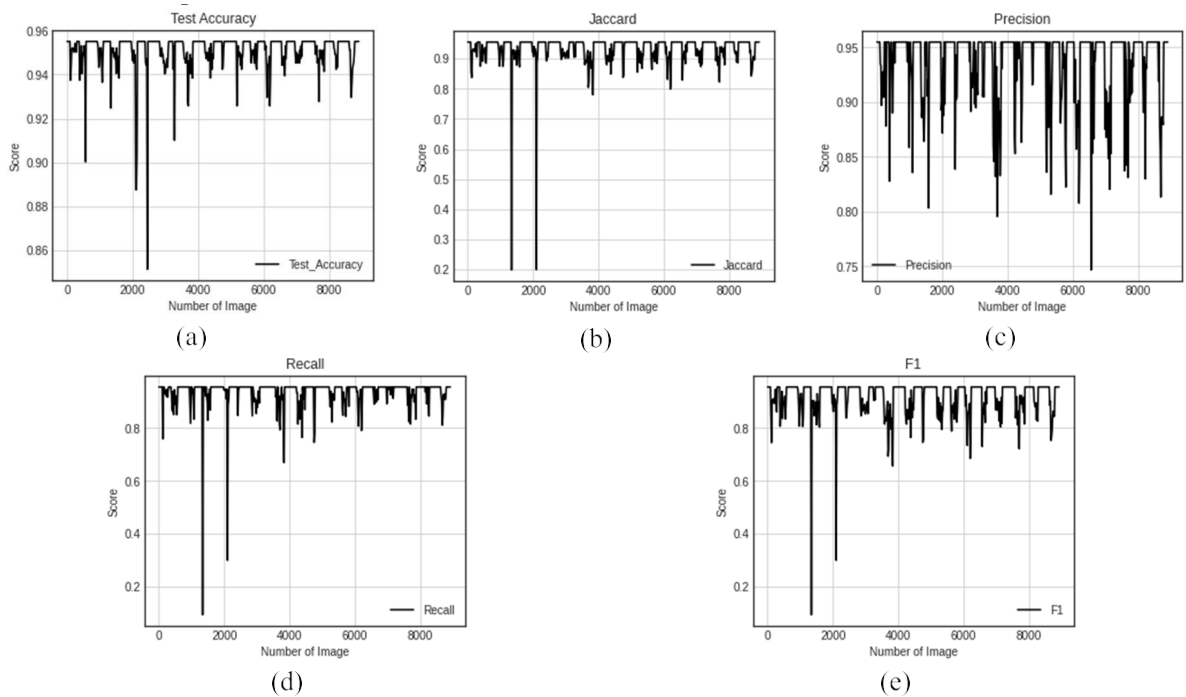

**FIGURE 12.** Testing accuracy measurement on test dataset COMBO_3, a) Accuracy, b) Jaccard, c) Precision, d) Recall, e) F1.

### D. PERFORMANCE MEASUREMENT ON 5-PATIENTS DATASET
#### 1) TRAINING ACCURACY MEASUREMENT
The modified UNet model was trained using five mediastinal LN slices from the TCIA dataset; each contained approximately 436 images on average. 10% of images from the training set were employed for tuning the model's learning performance. The model was trained for 100 epochs, and the estimated parameter's value is mentioned in Figure 13.

Figure 13 (a) depicts the model's training accuracy, and loss on the 5-patients dataset, 13 (b) illustrates validation accuracy and loss, 13 (c) shows intersect of union (IOU) score, 13 (d) displays dice coefficient score and 13 (e) mentions the score of recall. The model maintained 97.9% training accuracy with 94.1% validation. The training and validation loss was enormous at the origin of the training process. The losses were minimized fast with the increasing number of epochs concluding that the model could learn faster. The IOU, dice coefficient, and recall scores were 92.8%, 91.6%, and 95.7%.

#### 2) TESTING ACCURACY MEASUREMENT
##### a: EXPERIMENTAL ANALYSIS ON COMBO_1
Figure 14 (a) shows the value of the accuracy parameter, 14 (b) depicts the Jaccard score, 14 (c) illustrates precision, 14 (d) shows recall, and 14 (e) displays the F1 parameter's score. The model reached 93.2% accuracy, 89.4% Jaccard, 91.3% precision, 92.9% recall, and 92.0% F1.

##### b: EXPERIMENTAL ANALYSIS ON COMBO_2
Figure 15 (a) depicts the value of the accuracy parameter, 15 (b) displays the Jaccard score, 15 (c) represents precision, 15 (d) indicates recall, and 15 (e) illustrates the F1 parameter's score. The model achieved 94.1% accuracy, 90.6% Jaccard, 91.6% precision, 93.4% recall and 92.5% F1 accordingly.

##### c: EXPERIMENTAL ANALYSIS ON COMBO_3
Figure 16 (a) depicts the value of the accuracy parameter, 16 (b) illustrates the Jaccard score, 16 (c) shows precision, 16 (d) demonstrates recall, and 16 (e) displays the F1 parameter's score. The model gained 94.8% accuracy, 91.9% Jaccard, 93.1% precision, 94.1% recall, and 93.5% F1. The best performance was accomplished on this combination.

### E. MEASUREMENT SUMMARY
The proposed model performed better on the 5-patients dataset and COMBO_3 test dataset. Those dataset's mean and standard deviation were evaluated to investigate the actual reason. For TCIA, ELCAP, and 5-patient, we discovered 0.1881, 2085, and 0.2272 standard deviation scores and 0.2207, 0.4070, and 0.2226 mean scores. The standard deviation measures observed values or data deviate from the mean. When the standard deviation is close to zero, the measured values converge toward the mean. However, when the standard deviation is high, the values or data are dispersed and far from the mean. We can also say that all the data is equal when the standard deviation is 0. It means that a higher standard value denotes a sharp contrast image and sharp contrast emphasizes the comparison of several pixel colors. The model can learn different pixel values more accurately on high-contrast image datasets that directly impact the model's performance. The standard deviation of the 5-patients dataset is higher than other datasets, leading the proposed model to perform better. Though the proposed architecture gained the best accuracy on the 5-patients dataset, it also performed better on the other two. The model has maintained 91+% accuracy and 90+% precision score on both TCIA and ELCAP, which demonstrates the model's capability in segmenting and detecting LNs from different dataset images.

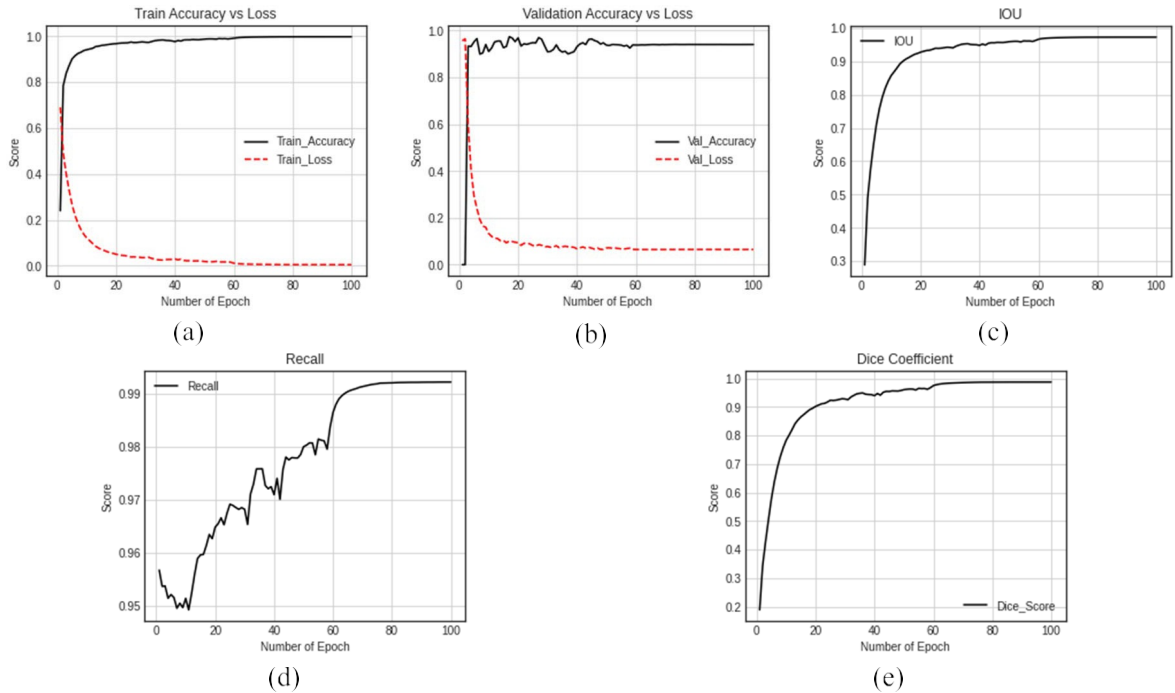

**FIGURE 13.** Training score evaluation a) Train Accuracy vs. Loss, b) Validation Accuracy vs. Loss, c) Intersect of Union (IOU), d) Dice Coefficient, e) Recall.

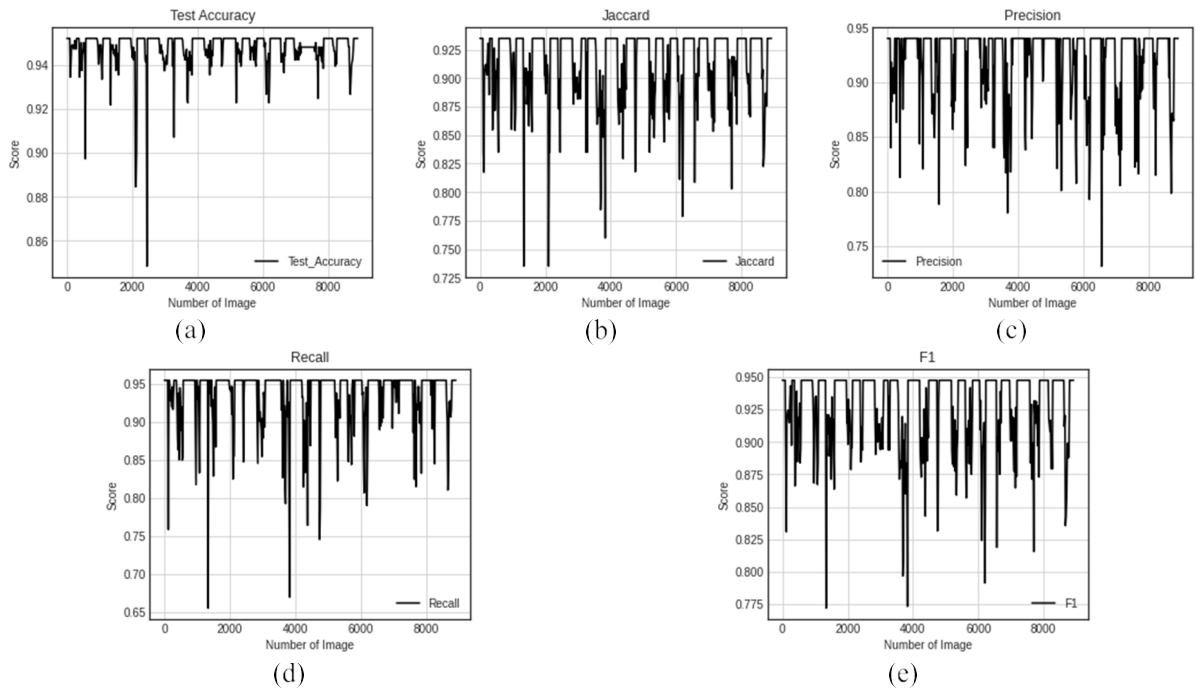

**FIGURE 14.** Testing accuracy measurement on test dataset COMBO_1, a) Accuracy, b) Jaccard, c) Precision, d) Recall, e) F1.

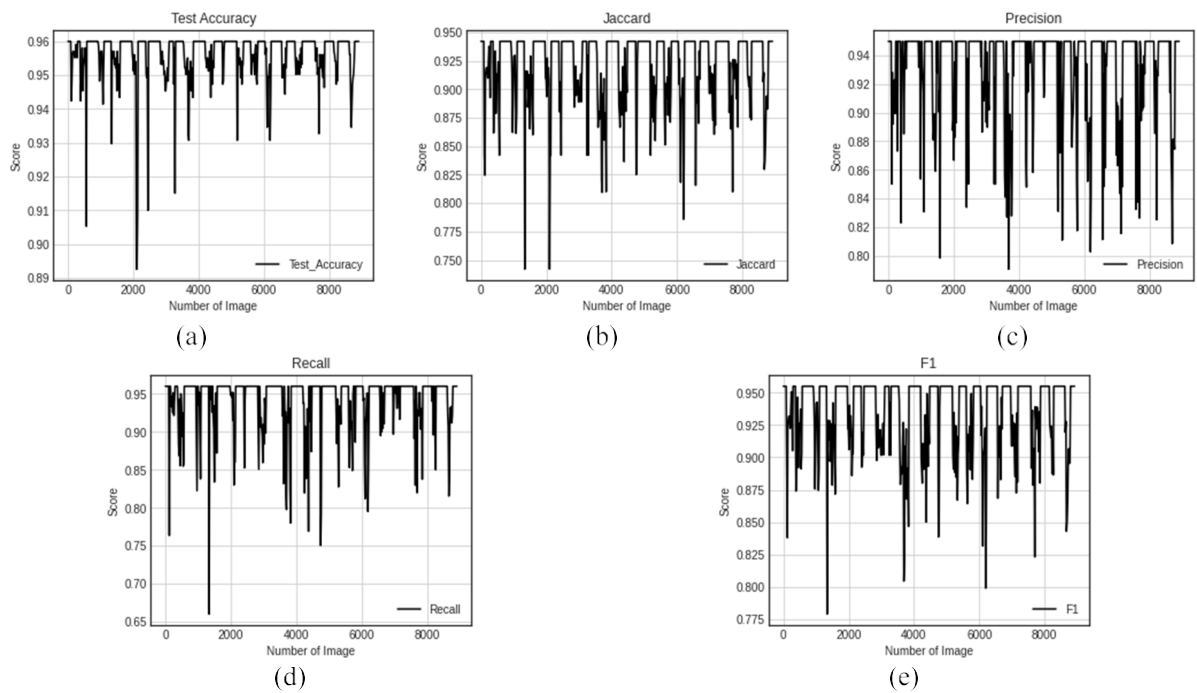

**FIGURE 15.** Testing accuracy measurement on test dataset COMBO_2, a) Accuracy, b) Jaccard, c) Precision, d) Recall, e) F1.

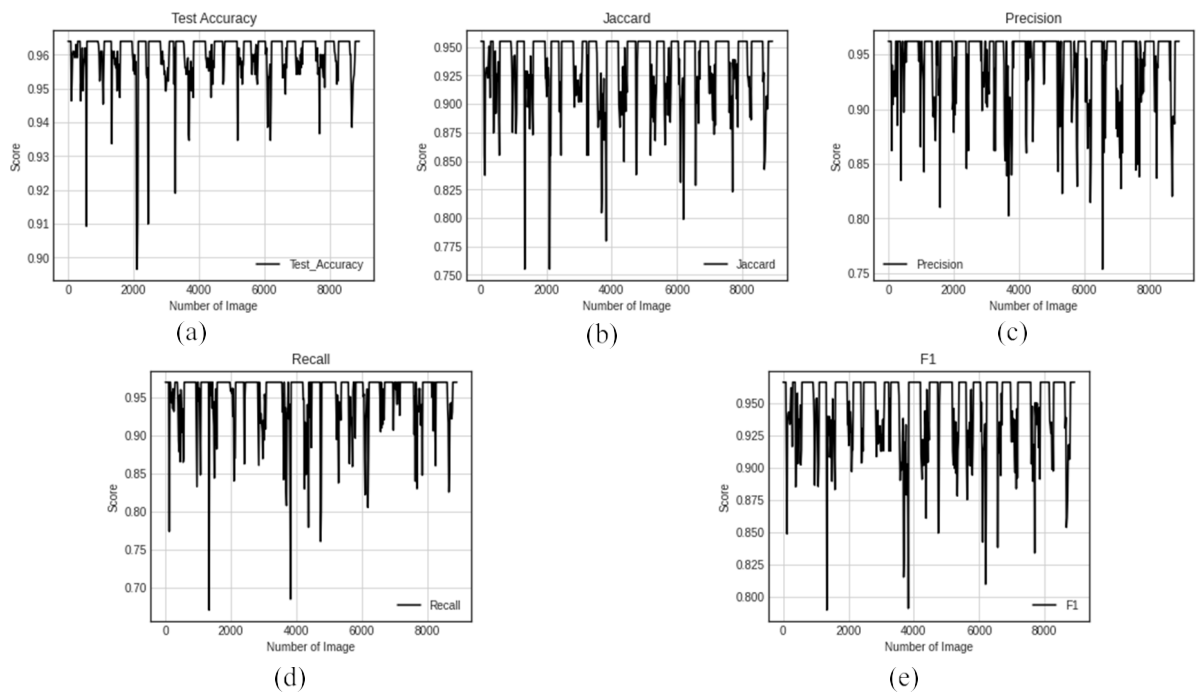

**FIGURE 16.** Testing accuracy measurement on test dataset COMBO_3, a) Accuracy, b) Jaccard, c) Precision, d) Recall, e) F1.

### F. PERFORMANCE COMPARISON
The model's performance was compared with other state-of-the-art techniques such as U-Net (without modification), AlexNet, SegNet, and Auto-LNDS (Mask RCNN-based Auto Lymph Node Detection and Segmentation, a proposed model by Zhao *et al.* [26]). The models were trained using TCIA, ELCAP, and 5-patients dataset and tensed using COMBO_1, COMBO_2, and COMBO_3 test datasets. The model's parameter values were kept similar, as mentioned in table 1.

TABLE 2. Old_U-Net model's performance.

| Dataset | Combo | FP/Vol | Accuracy | Dice | F1 | Recall | Precision |
|---|---|---|---|---|---|---|---|
| TCIA | 1 | 3.7 | 82.4% | 81.7% | 74.7 | 74.6% | 74.9% |
|  | 2 | 3.09 | 82.2% | 75.6% | 78.1 | 74.5% | **82.1%** |
|  | 3 | 2.7 | **84.9%** | 76.7% | 72.6 | 66.6% | 79.7% |
| ELCAP | 1 | 3.0 | 86.1% | 79.5% | 81.7 | 79.8% | 83.7% |
|  | 2 | 2.4 | 85.9% | 80.7% | 77.5 | 69.1% | **88.3%** |
|  | 3 | 2.9 | **86.7%** | 79.1% | 78.4 | 72.5% | 85.4% |
| 5-Patients | 1 | 2.5 | 87.02% | 81.7% | 75.3 | 75.5% | 75.1% |
|  | 2 | 2.48 | 87.9% | 77.5% | 79.1 | 72.6% | 86.9% |
|  | 3 | 2.36 | **89.01%** | 79.2% | 84.3 | 80.8% | **88.2%** |

1) PERFORMANCE MEASUREMENT USING Old_U-Net

In this research, the UNet++ model was altered with new upsampling techniques, and the implementation of the modified approach is explained in the previous section. Another experimental analysis was accomplished by employing the default transpose convolution for upsampling to distinguish the performance of altered UNet++ and original U-Net. No dense layer and no TGV-based technique were operated here. The model's performance was estimated on three different datasets and three test combinations, as mentioned in table 2.

According to table 2, Old_U-Net acquired the highest accuracy on the 5-patients training dataset and COMBO_3 test dataset. Still, its highest precision was developed on the ELCAP training dataset and COMBO_2 test dataset. Compared with the AlexNet, SegNet, and Auto-LNDS model's performance, the Old_U-Net scored the highest precision, but its accuracy was lower than that of AlexNet and SegNet. The model's overall performance maintains its second position among the four models.

2) PERFORMANCE MEASUREMENT USING ALEXNET

AlexNet is made up of eight layers, five of which are convolutional and three of which are completely connected. Rectified linear units (ReLU) and overlapping pooling are two significant elements of the AlexNet [27]. The ReLU nonlinearity allows significantly faster training than the normal tanh function. The overlapping pooling reduces training error and produces neighboring clusters of neurons without overlapping [28]. The network was trained using TCIA, ELCAP, and 5-patients dataset separately and tested using the test dataset combinations. The performance is mentioned in table 3.

AlexNet gained the highest accuracy and precision on the 5-patients training dataset and COMBO_3 test dataset mentioned in table 3. The model's highest accuracy outperformed Auto-LNDS, SegNet, and Old_U-Net, but its precision was lower than Auto-LNDS and Old_U-Net. The model's overall performance was better than SegNet, and among all models,

AlexNet executed better precision accuracy trained on the TCIA dataset.

3) PERFORMANCE MEASUREMENT USING SegNet

SegNet consists of an encoder and decoder network, followed by a pixel-wise classification layer. The encoder network has 13 convolutional layers that match with the first 13 convolutional layers of the VGG16 object classification network [29]. To minimize the number of parameters in the SegNet encoder network, we removed the ultimately linked layers to preserve higher resolution feature maps at the deepest encoder output. Each encoder layer has a corresponding decoder layer, with 13 layers in the decoder network. The output of the final decoder is sent into a multi-class softmax classifier, which generates class probabilities for each pixel separately. To generate a collection of feature maps, each encoder in the encoder network executes convolution with a filter bank. Then a rectified-linear nonlinearity (ReLU) is applied to each element [30]. The network was trained using TCIA, ELCAP, and 5-patients dataset separately and tested using the test dataset combinations. The performance is mentioned in table 4.

SegNet achieved the highest accuracy and precision on the 5-patient training and COMBO_3 test dataset shown in table 4. The model performed poorly than AlexNet, Old_U-Net, and Auto-LNDS, but the model's precision on the ELCAP training dataset was more reasonable than Auto-LNDS and AlexNet.

4) PERFORMANCE MEASUREMENT USING AUTO-LNDS (MODIFIED MASK RCNN)

For Auto-LNDS, the Mask RCNN is employed as the leading network. Mask R-CNN [31] is a system that can efficiently recognize objects in an image while producing a high-quality segmentation mask for each instance. The Feature Pyramid Network (FPN) [32], the Region Proposal Network (RPN) [33], and the head network Mask R-CNN create the Auto-LNDS. The backbone of Mask R-CNN was chosen to be Resnet-101 [34] to overcome the degradation problem and make it possible to train the deeper network. We used Mask R-CNN for nodal recognition and segmentation because the quantity and size of LNs vary from patient to patient. The network was trained using TCIA, ELCAP, and 5-patients dataset

TABLE 3. AlexNet model's performance.

| Dataset | Combo | FP/Vol | Accuracy | Dice | F1 | Recall | Precision |
|---|---|---|---|---|---|---|---|
| TCIA | 1 | 2.37 | 82.1% | 72.6% | 74.1% | 73.5% | 74.7% |
|  | 2 | 2.77 | 83.9% | 71.3% | 78.3% | 78.7% | 77.9% |
|  | 3 | 2.39 | **89.7%** | 82.2% | 81.5% | 80.7% | **82.3%** |
| ELCAP | 1 | 2.41 | 87.3% | 80.06% | 81.3% | 83.7% | 79.03% |
|  | 2 | 2.38 | **89.01%** | 83.03% | 79.9% | 81.9% | 78.01% |
|  | 3 | 2.25 | 88.5% | 80.2% | 81.05% | 82.1% | **80.04%** |
| 5-Patients | 1 | 2.43 | 88.4% | 84.04% | 82.1% | 82.2% | 82.05% |
|  | 2 | 2.20 | 89.03% | 83.08 | 81.05% | 79.3% | 82.9% |
|  | 3 | 2.32 | **90.01%** | 84.2% | 84.02% | 84.0% | **84.03%** |

TABLE 4. SegNet model's performance.

| Dataset | Combo | FP/Vol | Accuracy | Dice | F1 | Recall | Precision |
|---|---|---|---|---|---|---|---|
| TCIA | 1 | 2.43 | 82.8% | 72.4% | 75.7% | 75.7% | **75.7%** |
|  | 2 | 2.50 | 86.5% | 74.9% | 78.3% | 87.1% | 71.1% |
|  | 3 | 2.71 | **87.6%** | 77.6% | 76.4% | 73.6% | 79.4% |
| ELCAP | 1 | 3.02 | 86.01% | 74.1% | 79.01% | 77.1% | 81.06% |
|  | 2 | 2.78 | 88.03% | 75.02% | 77.04% | 73.4% | 81.05% |
|  | 3 | 3.00 | **88.5%** | 76.1% | 78.02% | 74.4% | **82.03%** |
| 5-Patients | 1 | 2.56 | 87.05% | 75.1% | 78.03% | 75.2% | 81.09% |
|  | 2 | 2.39 | 89.01% | 77.05% | 77.02% | 71.8% | 83.08% |
|  | 3 | 2.28 | **89.8%** | 78.02% | 80.01% | 76.3% | **84.00%** |

TABLE 5. Auto-LNDS model's performance.

| Dataset | Combo | FP/Vol | Accuracy | Dice | F1 | Recall | Precision |
|---|---|---|---|---|---|---|---|
| TCIA | 1 | 3.01 | **89.4%** | 85.7% | 81.0% | 81.5% | 80.5% |
|  | 2 | 3.75 | 85.9% | 80.1% | 82.6% | 84.9% | 80.4% |
|  | 3 | 2.62 | 87.4% | 84.7% | 80.1% | 77.5% | **82.9%** |
| ELCAP | 1 | 3.11 | 87.03% | 75.08% | 76.01% | 78.1% | 74.03% |
|  | 2 | 2.90 | **88.02%** | 77.3% | 79.01% | 78.9% | 79.1% |
|  | 3 | 2.73 | 87.1% | 74.7% | 77.9% | 77.7% | **78.06%** |
| 5-Patients | 1 | 2.33 | 87.5% | 76.09% | 78.02% | 76.1% | 80.07% |
|  | 2 | 2.89 | 87.9% | 78.03% | 78.05% | 74.4% | 82.02% |
|  | 3 | 2.40 | **89.02%** | 78.9% | 81.03% | 77.4% | **85.05%** |

separately and tested using the test dataset combinations. The performance is mentioned in table 5.

The Auto-LNDS model surpassed itself, having greater accuracy values using the TCIA dataset and COMBO_1 test dataset, as shown in Table 5. The model, learned from the images of the 5-patients dataset and tested on COMBO_3, had a better precision score than the score acquired from the TCIA and ELCAP datasets. The model's highest precision was 85.05% which was more significant than AlexNet and SegNet but lower than Old_U-Net. On the other hand, its accuracy score was third among those four models.

### G. DETECTION, SEGMENTATION, AND PRECISION CAPABILITY ANALYSIS

Table 6 displays some samples of different model's performance in segmenting LNs. Medical professionals reviewed the segmented images, and in their opinion, every model segmented the LN portions correctly from the first to third rows. In the fourth row, SegNet and the proposed model segmented accurately where Old_U-Net, AlexNet, and Auto-LNDS skipped one LN portion. The potential reason can be the tiny shape of that portion, lower image resolution, or the closeness of two portions that are complicated to disentangle. Similar issues were noticed in the fifth row, but the Auto-LNDS and proposed model performed better this time. Excluding Old_U-Net and SegNet, other models performed appropriate LN segmentation in row six. In row seven, only the proposed model recognized the tiny LN portion, which other models overlooked due to the small size, iso-intensity, or partial volume effect. All models correctly identified the LN portions in rows eight and nine. A similar problem as row seven was noticed in row ten, where the proposed model segmented the LN portion correctly. According to medical experts, the proposed method correctly segmented slight or closely situated LN portions.

Table 7 illustrates some specimens of different model's execution in detecting LNs. The specialists inspected the detected images. In their view, from the fifth column, the proposed model accurately detected LNs in both big and small sizes. Due to the slim size, the Auto-LNDS, Old_U-Net, SegNet, and AlexNet models skipped one LN in the second row. Due to poor image resolution and inconspicuous display on the LN image, one LN was missed by AlexNet and SegNet in row four. Two LNs were ignored by Old_U-Net, AlexNet, and SegNet in row five due to insufficient image resolution or partial volume effects. According to medical professionals, the proposed model's performance can be regarded as advanced and sounder than others in detecting LN portion influenced by the tiny shape, size, inadequate resolution, or partial volume effect.

**TABLE 6.** Lymph node segmentation and prediction by several models.

| | | | | Precision | | |
|---|---|---|---|---|---|---|
| CT Image | Actual | Old_U-Net | AlexNet | SegNet | Auto-LNDS | Propsoed Model |

**TABLE 7.** Lymph node detection by different models.

| CT Image | Old_U-Net | AlexNet | SegNet | Auto-LNDS | Proposed Model |
|---|---|---|---|---|---|
| 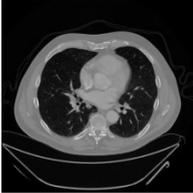 | 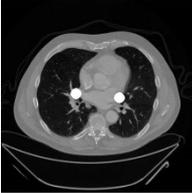 | 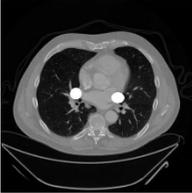 | 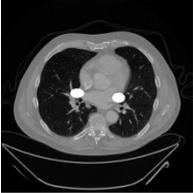 | 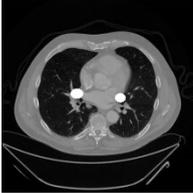 | 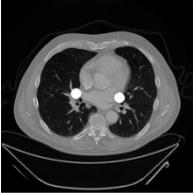 |
| 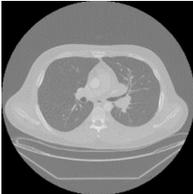 | 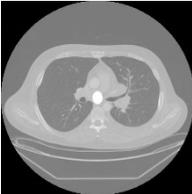 | 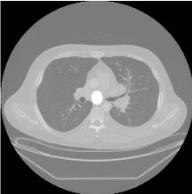 | 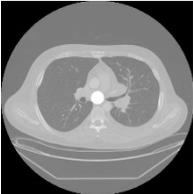 | 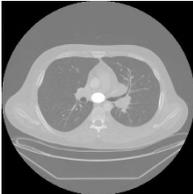 | 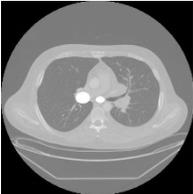 |
| 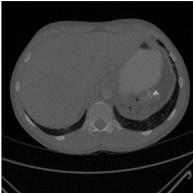 | 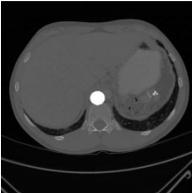 | 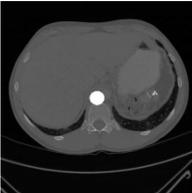 | 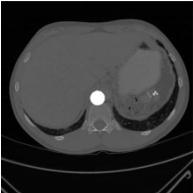 | 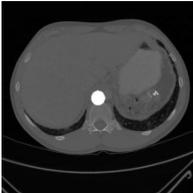 | 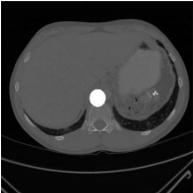 |
| 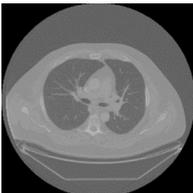 | 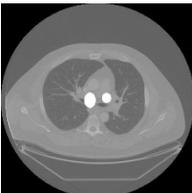 | 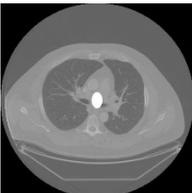 | 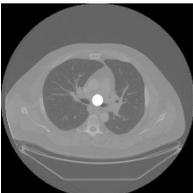 | 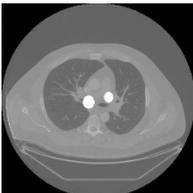 | 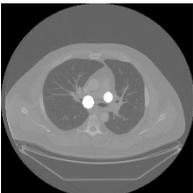 |
| 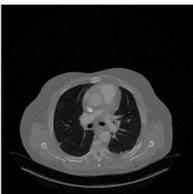 | 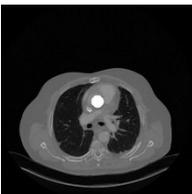 | 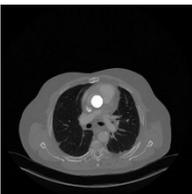 | 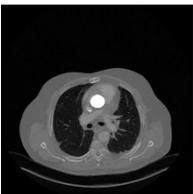 | 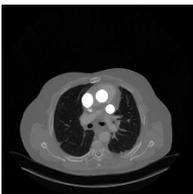 | 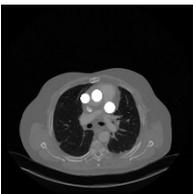 |

## V. DISCUSSION & LIMITATION

Detecting all the LNs is the initial stage in the treatment process, and it is a tedious and time-consuming task. We have designed a novel deep learning procedure for rapid and reliable detection and segmentation of mediastinal LNs. This technology allows radiologists to quickly acquire an LN map within a few seconds to aid in diagnostic interpretation. It is the first attempt at autonomously detecting and segmenting LNs based on CT data. Previous attempts with various approaches concentrated on contrast-enhanced or small diameter LNs. Besides, previous approaches performed poorly while dealing with CT images containing tiny LN portions. Many approaches overlooked the small LN portion in segmentation or overlapped in detection. We have given considerable importance to the minor portion to minimize information loss during training. The hybridized bilinear interpolation method performed information lossless upsampling, and the TGV-based technique smoothed the image by removing noise particles. The processed datasets with sharp contrast and quality contributed remarkably to the accuracy. The model achieved a maximum of 94.8% accuracy, 91.9% Jaccard, 94.1% recall, and 93.1% precision score, which was better than state-of-the-art strategies.

This research contains several limitations. We focused on detecting and segmenting mediastinal Lymph Nodes from CT images. A modified deep learning approach UNet++

was used to solve the minor portion detection issues of medical images. The objective of the proposed method was to maintain texture discontinuities, select noisy areas, search for appropriate balance points through backpropagation, and recreate image resolution. High-resolution CT images were employed for training and testing purposes, and contrast-enhanced CT images were avoided. Considering the time limitation, three datasets were analyzed, and the model's accuracy was compared with four approaches. Still, more public datasets and approaches are available. Detection and segmentation quality were prioritized more in this research, and the model may not perform in real-time LN detection.

## VI. CONCLUSION

Lymph node detection plays a crucial function in detecting cancer and assures potential treatment opportunities for the patients. A modified UNet++ model has been presented in this article. The UNet+ was modified using bilinear interpolation and total generalized variation (TGV) based upsampling technique for segmenting and detecting mediastinal lymph nodes from CT images. Receiving assistance from medical professionals, a dataset with three different combinations was processed. The modified UNet model's performance was calculated and compared with state-of-the-art strategies. The model gained the highest 94.8% accuracy and 93.1% precision accuracy with a 4.7 false-positive rate per volume which was more promising than other techniques. The model's performance was estimated on various datasets, and a satisfactory result was achieved. It can be concluded that the UNet+ model with the offered strategy can detect and segment mediastinal lymph nodes accurately. In the future, we intend to accomplish several filter modifications to decrease the image's noise and improve performance.

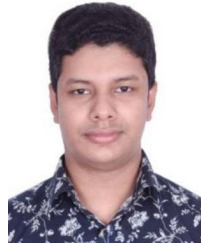

AL-AKHIR NAYAN received the Bachelor of Science degree in computer science and engineering from the University of Liberal Arts Bangladesh (ULAB), Dhaka, Bangladesh, in 2019. He is currently pursuing the master's degree with the Department of Computer Engineering, Chulalongkorn University, Bangkok, Thailand. He joined the European University of Bangladesh (EUB), Dhaka, in 2019, and worked as a Lecturer with the Department of Computer Science and Engineering. His research interests include deep learning, machine learning, artificial intelligence, medical image processing, and the IoT.

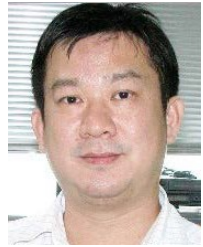

BOONSERM KIJSIRIKUL received the B.Eng. degree in electronic and electrical engineering, the M.Sc. degree in computer science, and the Ph.D. degree in computer science from the Tokyo Institute of Technology, Japan, in 1986, 1990, and 1993, respectively. He is currently a Professor at the Department of Computer Engineering, Chulalongkorn University, Thailand. His current research interests include machine learning, artificial intelligence, and natural language processing.

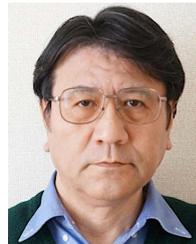

YUJI IWAHORI (Member, IEEE) received the B.S. degree from the Nagoya Institute of Technology, in 1983, and the M.S. and Ph.D. degrees from the Department of Electrical and Electronics, Tokyo Institute of Technology, in 1985 and 1988, respectively. He joined the Educational Centre for Information Processing, Nagoya Institute of Technology, as a Research Associate, in 1988, where he became a Professor with the Centre for Information and Media Studies, in 2002. He has been a Visiting Researcher at UBC Computer Science, since 1991. He has been with Chubu University as a Professor, since 2004, where he acted as the Department Head of Computer Science, the Head of Graduate Course of Computer Science, and the Vice-Dean of the College of Engineering. He has been a Research Collaborator with IIT Guwahati, since 2010, and with the Department of Computer Engineering, Chulalongkorn University, since 2014. He has become an Honorary Faculty at IIT Guwahati, since 2020. His research interests include computer vision, biomedical image processing, deep learning, and application of artificial intelligence. He got the KES 2008 Best Paper Award and the KES 2013 Best Paper Award from KES International.

● ● ●